\documentclass[aps,twocolumn,showpacs]{revtex4}
\usepackage{graphicx}
\begin{document}
\title{Study on the One-Proton Halo Structure in $^{23}$Al
\footnote{\scriptsize Supported by the Major State Basic
Research Development Program in China Under Contract No. G2000774004,
the National Natural Science Foundation of China (NNSFC) under Grant
No 10405032, 10328259 and 10135030 and the Phosphor program in Shanghai under
contract No. 03QA14066.}}
\author{FANG De-Qing$^{1,}$
\footnote{\scriptsize Corresponding author.
{\it Email address: fangdq@sinr.ac.cn}}}
\author{MA Chun-wang$^{1,2}$}
\author{MA Yu-Gang$^1$}
\author{CAI Xiang-Zhou$^1$}
\author{CHEN Jin-Gen$^{1,2}$}
\author{CHEN Jin-hui$^{1,2}$}
\author{GUO Wei$^1$}
\author{TIAN Wen-Dong$^1$}
\author{WANG Kun$^{1,2}$}
\author{WEI Yi-Bin$^{1,2}$}
\author{YAN Ting-Zhi$^{1,2}$}
\author{ZHONG Chen$^1$}
\author{ZUO Jia-xu$^{1,2}$}
\author{SHEN Wen-Qing$^1$}

\affiliation{\footnotesize $^1$Shanghai Institute of Applied
Physics, Chinese Academy of Sciences, P. O. Box 800-204, Shanghai
201800}
\affiliation{\footnotesize $^2$Graduate school of Chinese
Academy of Sciences}
\date{\today}

\begin{abstract}
The Glauber theory has been used to investigate the reaction cross
section of proton-rich nucleus $^{23}$Al. A core plus a proton
structure is assumed for $^{23}$Al. HO-type density distribution
is used for the core while the density distribution for the
valence proton is calculated by solving the eigenvalue problem of
Woods-Saxon potential. The transparency function in an analytical
expression is obtained adopting multi-Gaussian expansion for the
density distribution. Coulomb correction and finite-range
interaction are introduced. This modified Glauber model is apt for
halo nuclei. A dominate $s$-wave is suggested for the last proton
in $^{23}$Al from our analysis which is possible in the RMF
calculation.
\end{abstract}

\maketitle

Interest in reaction cross section ($\sigma_R$) has grown over the
last decades with the development of radioactive ion beams (RIBs).
$\sigma_R$ for light exotic nuclei has been studied extensively
both theoretically and experimentally. Up to now much effort has
been devoted to the study of neutron halo structure, such as the
discovery of neutron halo nuclei $^{6,8}$He, $^{11}$Li,
$^{11,14}$Be \cite{TAN} and the recent work on $^{19}$C
\cite{BAZ,OZA1}. At the same time, some work has been performed on
proton halo nuclei, e.g. $^{8}$B and $^{17}$Ne. Due to Coulomb
effect, the formation of proton halo is more difficult and
complicated compared to neutron halo structure. And discrepancies
were found in proton halo nuclei, especially on $^{8}$B
\cite{WAR1,NEG,FUK,OBU}. Recently the $\sigma_R$ was measured for
the proton halo candidate $^{23}$Al, whose separation energy of
the last proton is very small (0.125 MeV) \cite{AUD}. Enhancement
in $\sigma_R$ was observed for $^{23}$Al compared to its neighbors
\cite{CAI,ZHA}. According to the shell model, the last proton in
$^{23}$Al is in the 1$d_{5/2}$ orbit. But the RMF calculation
shows that it is possible to have the inversion of the 1$s_{1/2}$
and 1$d_{5/2}$ orbit in $^{23}$Al \cite{CHE}. Thus it is very
interesting and important to know which orbit the last proton
occupies.

Several methods are available to  study the total reaction cross
section, i.e. the  multi-step scattering theory of Glauber
\cite{GLA}, the transport model method of Ma et al. \cite{Ma} and
the semi-empirical formula of Kox et al. \cite{Kox} and Shen et
al. \cite{Shen} etc. For the first one, the optical limit
approximation of the Glauber approach  is the most common used
method \cite{GLA}. In the high energy domain, the Glauber model is
based on the individual nucleon-nucleon collisions along the
classical straight-line trajectory in the overlap volume of the
colliding nuclei \cite{GLA,KAR}. This model has been extended to
low energies by taking into account the Coulomb distortion of the
eikonal trajectory \cite{CHA}. Recently, it has been found that
the Glauber model always underestimates $\sigma_R$ by 10-50$\%$ at
intermediate energies, if one assumes HO-type nucleon density
distributions and determines the width parameter by reproducing
the interaction cross section at relativistic energies
\cite{OZA2}. To improve the calculation of the Glauber model at
low energies, the finite-range interaction was considered and its
energy dependence was investigated by fitting the $\sigma_R$ of
$^{12}$C + $^{12}$C from low to high energies \cite{ZHE}. And also
the few-body Glauber model was widely used to study nuclei with
halo structure, which assumes core plus one-(two-)nucleon
structure \cite{OGA}. However, all these calculations are done by
solving a multi-dimensional numerical integration or Monte-Carlo
simulation. To simplify the calculation, the Glauber model in an
analytical form has been obtained if we use Gaussian function for
density distributions \cite{CHA}. But this method can only be used
to describe stable nuclei. In order to study nuclear reaction
involving exotic nuclei, the separation energy dependent
diffuseness was introduce into this method \cite{FEN}. Here we
present a more general analytical form by expanding the density
distributions of projectile and target using multi-Gaussian
functions, which can be used to calculate $\sigma_R$ for both
stable nuclei and nuclei with halo structure.

To begin with, we describe the formulation of $\sigma_R$ in the framework
of Glauber model in the optical limit approximation. If we distinguish
neutrons and protons in the projectile and target, the reaction cross
section can be expressed as
\begin{equation}
\sigma_{\mbox{\scriptsize R}}=2\pi\int_0^{\infty} [1-T(b)]b\mbox{d}b
\end{equation}
\noindent
where $T(b)$ is the transparency function at impact parameter $b$
\begin{eqnarray}
T(b)& = & \exp\{-\int dr\int f(r-r')\sum_{\mbox{i=n,p}} \nonumber\\
&& \sum_{\mbox{j=n,p}}\sigma_{\mbox{ij}}\rho^z_{\mbox{\scriptsize Pi}}(r)
\rho^z_{\mbox{\scriptsize Tj}}(r'-b)dr'\}
\end{eqnarray}
here $f(r)$ is the finite-range function
\begin{equation}
\begin{array}{c}
f(r)=\frac{1}{\pi\gamma_0^2}\exp(-\frac{r^2}{\gamma_0^2})
\end{array}
\end{equation}
with $\gamma_0$ being the finite-range parameter.
$\sigma_{\mbox{ij}}$ is the nucleon-nucleon collision cross
section. $\rho^z_{\mbox{\scriptsize ij}}$ is the neutron (proton)
density distribution in the projectile (target) with the z axis
integrated \cite{CHA}. With the input density distribution of the
projectile and target, the reaction cross section can be obtained
after the multi-dimensional numerical integration.

If we assume Gaussian distributions for the density of projectile
and target, which is suitable for light stable nuclei. Then we can
obtain an analytical transparency function \cite{CHA}. To extend
this method for both stable nuclei and halo nuclei with a long
tail in the density distribution, we express the density
distribution by N Gaussian functions
\begin{eqnarray}
\rho_{\mbox{\scriptsize ij}}(r)=\sum_{\mbox{\scriptsize k=1}}^{\mbox{\scriptsize N}}
\rho^{\mbox{\scriptsize k}}_{\mbox{\scriptsize ij}}(0)
e^{-(r/a^{\mbox{\scriptsize k}}_{\mbox{\scriptsize ij}})^2} \hspace{.5cm} (\mbox{i=P,T; j=n,p})
\end{eqnarray}
where $\rho^{\mbox{\scriptsize k}}_{\mbox{\scriptsize ij}}(0)$ and
$a^{\mbox{\scriptsize k}}_{\mbox{\scriptsize ij}}$ are the
amplitude and width parameters of the Gaussian functions. Then the
transparency function can be written in an analytical form
similarly \cite{CHA,FEN}
\begin{eqnarray}
T(b)& = & \exp\{ -\pi^2\sum_{\mbox{\scriptsize i=n,p}}\sum_{\mbox{\scriptsize j=n,p}}
\sum_{\mbox{\scriptsize k=1}}^{\mbox{\scriptsize N}}
\sum_{\mbox{\scriptsize k}^{\prime}=1}^{\mbox{\scriptsize N}^{\prime}}
\sigma_{\mbox{\scriptsize ij}}\nonumber\\
&&\times\rho^{\mbox{\scriptsize k}}_{\mbox{\scriptsize Pi}}(0)
\rho^{\mbox{\scriptsize k}^{\prime}}_{\mbox{\scriptsize Tj}}(0)
\frac{(a^{\mbox{\scriptsize k}}_{\mbox{\scriptsize Pi}})^3
(a^{\mbox{\scriptsize k}^{\prime}}_{\mbox{\scriptsize Tj}})^3}
{(a^{\mbox{\scriptsize k}}_{\mbox{\scriptsize Pi}})^2+
(a^{\mbox{\scriptsize k}^{\prime}}_{\mbox{\scriptsize Tj}})^2+
\gamma_0^2}\nonumber\\
&&\times\exp[-\frac{b'^2}{(a^{\mbox{\scriptsize k}}_{\mbox{\scriptsize Pi}})^2+
(a^{\mbox{\scriptsize k}^{\prime}}_{\mbox{\scriptsize Tj}})^2+\gamma_0^2}]\}
\end{eqnarray}
where $b'$ is the impact parameter after considering Coulomb correction
\begin{equation}
b'^2=\frac{b^2}{1-V_{\mbox{\scriptsize c}}/E_{\mbox{\scriptsize CM}}}
    =\frac{b^2}{1-1.44Z_{\mbox{\scriptsize P}}Z_{\mbox{\scriptsize T}}
    /(R_{\mbox{\scriptsize int}}E_{\mbox{\scriptsize CM}})}
\end{equation}
here $Z_{\mbox{\scriptsize P}}$($Z_{\mbox{\scriptsize T}}$) refers
to the charge number of the projectile (target),
$R_{\mbox{\scriptsize int}}$ is the interaction radius and
$E_{\mbox{\scriptsize CM}}$ is the incident energy in the center
of mass frame \cite{CHA}.

To investigate the $\sigma_R$ of $^{23}$Al, we assume a core
($^{22}$Mg) plus one proton structure since the weak bind between
the last proton and the core. HO-type distribution is used for the
core density. The density distribution for the valence proton was
calculated by solving the eigenvalue problem of Woods-Saxon
potential \cite{SPE}
\begin{equation}
\begin{array}{c}
\frac{d^2R(r)}{dr^2}+\frac{2\mu}{\hbar^2}[E-U(r)-\frac{l(l+1)\hbar^2}{2\mu r^2}]R(r)=0 \\
U(r)=-V_0f(r)+V_{ls}({\bf l\cdot s})r_0^2\frac{1}{r}\frac{d}{dr}f(r)+V_{\mbox{\scriptsize Coul}}
\end{array}
\end{equation}
where $f(r)=[1+\exp(\frac{r-R}{a})]^{-1}$ with $R=r_0A^{1/3}_{c}$ $(V_{ls}=17 \mbox{MeV})$.
$V_0$ is the depth of potential, $V_{\mbox{\scriptsize Coul}}$ is the Coulomb potential.
In the calculation the diffuseness ($a$) and radii parameter ($r_0$) were chosen
to be 0.67 fm and 1.27 fm\cite{YAM}.

First we assume HO-type distribution for the $^{12}$C target and
determine the HO width parameter by reproducing the interaction
cross section of $^{12}$C + $^{12}$C at relativistic energy. Then
the $\sigma_R$ for $^{12}$C + $^{12}$C at different energies are
calculated and compared them with experimental data. In the
calculation, the finite-range parameter was taken from Ref.
\cite{ZHE}.
\begin{equation}
\begin{array}{l}
\gamma_0^2=2\beta^2 \\
\beta=0.996\exp(-\frac{E}{106.679})+0.089
\end{array}
\end{equation}
where $E$ is the incident energy in the laboratory frame in MeV.
With the consideration of the energy dependent finite-range
interaction, the underestimation at low energies in the zero-range
Glauber model was removed as described in Ref. \cite{ZHE}

\begin{figure}[b]
\begin{center}
\includegraphics[width=8.9cm,angle=0.]{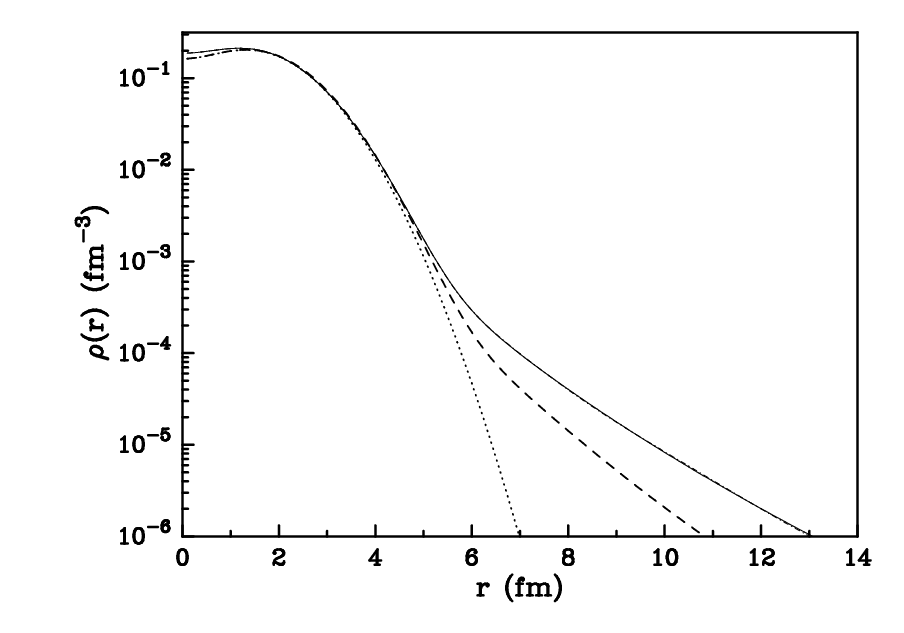}
\end{center}
\caption{Density distributions of $^{22}$Mg and $^{23}$Al. The
dotted line is the density distribution of $^{22}$Mg. The solid
and dashed line are the density distribution of $^{23}$Al in $s$-
and $d$-wave, respectively. For comparison, the density
distribution of $^{23}$Al in the $s$-wave calculated by
multi-Gaussian expansion (Eq.(4)) is also plotted as the
dotted-dashed line. It is almost overlapped with the original
density distribution.} \label{rho}
\end{figure}

\begin{table}[h]
\begin{center}
\caption{The reaction cross sections for $^{22}$Mg and $^{23}$Al with carbon target.}
\begin{tabular}{c c r l}
\hline
Nuclei & energy ($A$MeV) & $\sigma_{R}$ (mb) & reference \\
\hline
$^{22}$Mg & 33.4 &  1531$\pm$125  & \cite{CAI}\\
$^{23}$Al & 35.9 &  1892$\pm$145  & \cite{CAI}\\
\hline
\end{tabular}
\label{data}
\end{center}
\end{table}

Then we assume HO distribution for the $^{22}$Mg core, and adjust
the HO width parameter to fit the measured $\sigma_R$ at 33.4$A$
MeV as given in Table. \ref{data}.

Since spin parity assignment for the ground state of $^{23}$Al is unknown
experimentally, we took the assumed value(1/2$^+$) from Ref.\cite{ZHA}.
The mixing configuration for the last proton was considered as following
\begin{equation}
\begin{array}{rl}
\phi \propto & \sqrt{f} [^{22}\mbox{Mg}(0^+)\otimes\nu_{s_{1/2}}]_{J=1/2^+} \\
& +\sqrt{1-f} [^{22}\mbox{Mg}(2^+)\otimes\nu_{d_{5/2}}]_{J=1/2^+}
\end{array}
\end{equation}
where $f(0\le f\le1)$ denotes the relative $s$-wave spectroscopic
factor. $\nu_{s_{1/2}}(r)$ and $\nu_{d_{5/2}}$ refer to the
wavefunction of $s$- and $d$-wave. The density distribution of the
valence proton can be calculated by $\phi^2(r)$. The wavefunction
for the valence proton in 2$s_{1/2}$ or 1$d_{5/2}$ orbit was
calculated by adjusted the depth of Woods-Saxon potential to
reproduce the separation energy of the last proton in $^{23}$Al.
The density distribution of $^{23}$Al was obtained by adding that
of the core and the valence proton. For simplicity, we used the same
density distributions for both the ground($0^+$) and excited($2^+$) states
of $^{22}$Mg. Comparison of $^{23}$Al's
density distributions in $s$- and $d$-wave with the core was shown
in Fig.\ref{rho}. From the figure, the density distributions of
$^{23}$Al are larger than that of $^{22}$Mg. The tail of $s$-wave
is much longer than $d$-wave. In the expansion, we used 12
Gaussian functions for the projectile and 4 Gaussian functions for
the target density distribution. In the fitting, $\chi^2$ is less
than 10$^{-9}$. As we can see in Fig.\ref{rho}, good agreement
between the original and fitted density was obtained.

\begin{figure}[h]
\begin{center}
\includegraphics[width=8.5cm,angle=0.]{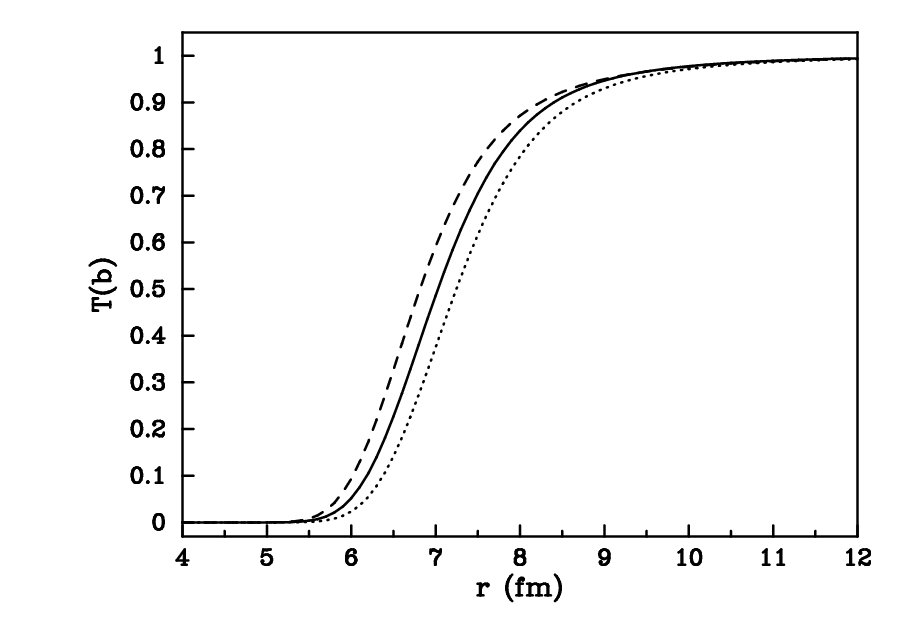}
\end{center}
\caption{Transparency function (T(b)) for $^{23}$Al in the $s$-wave on $^{12}$C
target at 35.9$A$ MeV. The dotted line is calculated by the Glauber model in
the optical limit approximation. The dashed line is the results with finite-range
interaction. And the solid line is calculation by the Glauber model with both
finite-range interaction and Coulomb correction.}
\label{tb}
\end{figure}

With the Gaussian parameters, we can directly calculate the
transparency function (T(b)) using Eq.(5). The effect of
finite-range and Coulomb correction on T(b) was compared in Fig.
\ref{tb}. As we can see, T(b) is saturated to unity as the impact
parameter increase. Compared with the Glauber model in the optical
limit approximation, the finite-range effect makes T(b) decrease
slower as the decrease of the impact parameter which leads to
larger reaction cross section. But the Coulomb correction makes
T(b) decrease faster as the decrease of the impact parameter.

\begin{figure}[b]
\begin{center}
\includegraphics[width=8.9cm,angle=-0.]{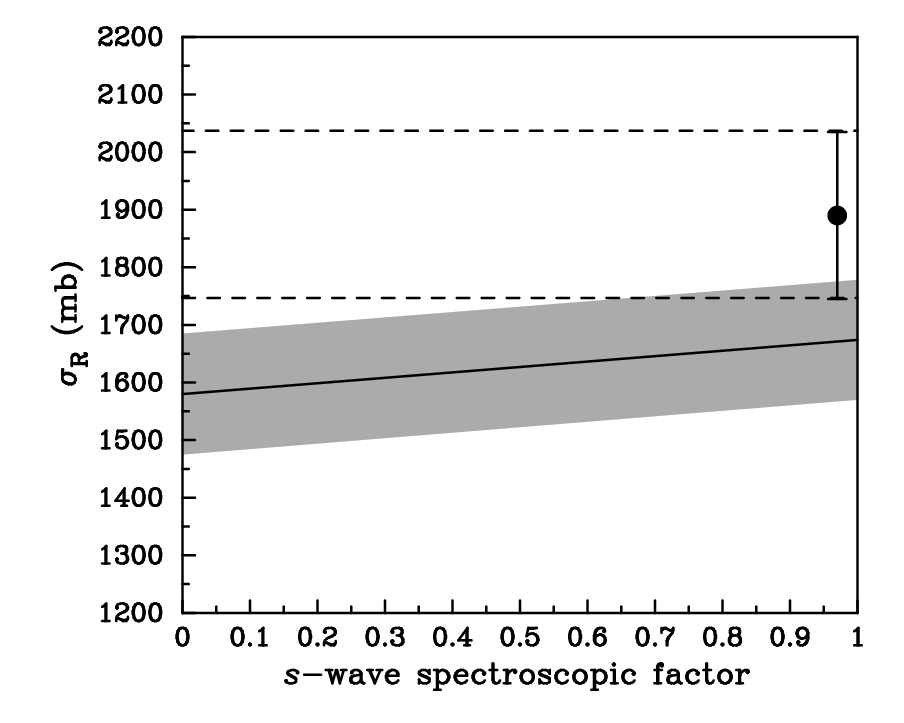}
\end{center}
\caption{Dependence of the reaction cross section on the $s$-wave spectroscopic
factor. The shaded area comes from the experimental error of $^{22}$Mg and
$^{23}$Al's $\sigma_R$ data. The dot is the $\sigma_R$ of $^{23}$Al.}
\label{sfactor}
\end{figure}

\begin{figure}[t]
\begin{center}
\includegraphics[width=8.9cm,angle=-0.]{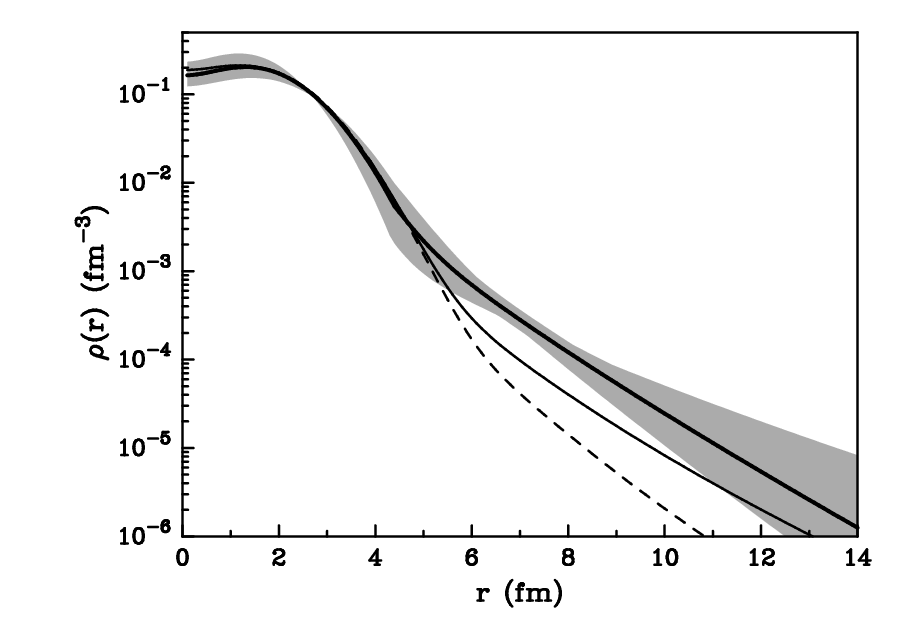}
\end{center}
\caption{Density distributions of $^{23}$Al. The solid and dashed
line are the density distribution of $^{23}$Al in $s$- and
$d$-wave, respectively. The thick line is the extracted density
distribution assuming HO plus Yukawa distribution. The shaded area
is the error from the experimental $\sigma_R$ data.}
\label{density}
\end{figure}

Dependence of the reaction cross section on the $s$-wave
spectroscopic factor was calculated and shown in Fig.
\ref{sfactor}. The $\sigma_R$ of $^{23}$Al was also plotted
\cite{CAI}. It is found that within the uncertainties our analysis
suggests dominate $s$-wave for the last proton in $^{23}$Al. The
lower limit of the $s$-wave spectroscopic factor is around 0.65.
But it should be pointed out that the error of $\sigma_R$ for
$^{23}$Al is comparable with the difference in $\sigma_R$ between
$s$- and $d$-wave. And even the $s$-wave calculation still underestimates
the experimental $\sigma_R$ data. Zhao et al. suggested an enlarged core
in $^{23}$Al which lead to larger reaction cross section\cite{ZHAO}.
That maybe one of the reasons for the underestimation in our calculations.
In order to obtain more confirmative conclusion, much more precise
measurement and theoretical study of $\sigma_R$ for
$^{23}$Al is required. And the measurement of momentum
distribution for the residue after one-proton remove is expected
which is much easy to distinguish $s$- and $d$-wave.

For comparison, we also assume HO core plus Yukawa-square tail for
the density distribution of $^{23}$Al and determine the parameters
by reproducing the experimental $\sigma_R$. The Yukawa-square tail
is known to be a good approximation to the shape of a
single-particle density at the outer region of a core with
centrifugal and Coulomb barriers \cite{FUK}. The assumed density
is written as
\begin{equation}
\rho (r) = \left \{ \begin{array}{lc}
\text{HO-type}  & (r<r_{\text{c}})  \\
Y\exp (-\lambda r)/r^2 & (r\ge r_{\text{c}})
\end{array} \right.
\label{rhoyuk}
\end{equation}
where $r_{\text{c}}$ is the crossing point of these two functions,
the factor $Y$ is to keep the equality of the two distributions at
$r_{\text{c}}$. The width parameter of the HO-type core is fitted
to the $\sigma_R$ of $^{22}$Mg. Then the parameters $\lambda$ and
$r_{\text{c}}$ are fitted to the  $\sigma_R$ of $^{23}$Al given in
Table. \ref{data} with an additional normalization condition
$\int\rho(r)d^3r=Z$ and $Z$ is the charge number of $^{23}$Al.

In Fig.\ref{density}, the extracted density distribution was compared with the
calculated density for $^{23}$Al in $s$- and $d$-wave. The extracted density
is closer to $s$-wave than $d$-wave. It demonstrates a long tail in the density
distribution and supports the conclusion that the $s$-wave is dominate for the
last proton in $^{23}$Al.

In summary, the Glauber theory has been used to investigate the reaction
cross section. Adopting multi-Gaussian expansion for the density distribution
of projectile and target, an analytical form for the transparency function
was deduced. The energy dependent finite-range interaction and Coulomb correction
were considered. It has been pointed out that this modified Glauber model is
suitable for studying the $\sigma_R$ of both stable nuclei and nuclei with
exotic structure. We used this model to study the reaction cross section of
proton-rich nucleus $^{23}$Al. A core plus proton structure was assumed for $^{23}$Al.
HO-type density distribution was used for the core while the density distribution
for the valence proton was calculated by solving the eigenvalue problem of
Woods-Saxon potential. The $s$- and $d$-wave mixing configuration was studied.
Density distribution was extracted for $^{23}$Al assuming HO plus Yukawa-square
tail. A dominate $s$-wave was suggested in our analysis which indicates a proton
halo structure in $^{23}$Al.

\end{document}